\documentstyle[12pt]{article}

\begin{document}
\vspace{10mm}
\hspace{80mm}  Mod. Phys. Lett. A 12(1997)2859.
\begin{center}
\bf    Relativistic semiclassical wave equation and its solution  \\
\vspace{5mm}
\rm                   {   M. N. Sergeenko   } \\
\vspace{2mm}
\it  { The National Academy of Sciences of Belarus, Institute of Physics,\\
                      Minsk 220072, Belarus \ and   \\
             Gomel State University, Gomel 246699, Belarus }
\end{center}

\begin{abstract} The properties of relativistic particles in the
quasiclassical region are investigated. The relativistic
semiclassical wave equation appropriate in the quasiclassical region
is derived. It is shown that the leading-order WKB quantization rule
is the appropriate method to solve the equation obtained.
\\ ~ \\
\noindent PACS number(s): 03.65.Ge, 03.65.Sq
\end{abstract}

\noindent {\bf 1. Introduction }\\

It is generally agreed that the full description of quantum systems
is given by quantum mechanics. However, investigations by many
authors show that, in fact, there exists an alternative approach.
This is the approach based on the quasiclassical method.

The general form of the semiclassical description of
quantum-mechanical systems has been considered in Ref. \cite{Mi}.  It
was shown that the semiclassical description resulting from Weyl's
association of operators to functions is identical with the quantum
description and no information need to be lost in going from one to
the another. The semiclassical approach merely becomes a different
representation of the same algebra as that of the quantum mechanical
system, and then the expectation values, dispersions, and dynamics of
both become identical. What is more "the semiclassical description is
more general than quantum mechanical description..." \cite{Mi}.

The quasiclassical method is based on the correspondence principle
between classical functions and operators of quantum mechanics and
with a real, but not necessary positive, probability density function
in phase space corresponding to a particular quantum-mechanical
state. The correspondence principle is used to derive the wave
equation in quantum mechanics. In Ref. \cite{Se} this principle has
been used to derive the semiclassical wave equation appropriate in
the quasiclassical region. It was shown that the standard WKB method
(to leading order in $\hbar$) is the appropriate method to solve this
equation. Solution of the obtained wave equation by this method
yields the exact eigenvalues for {\em all} solvable spherically
symmetric potentials. The corresponding eigenfunctions have the same
behavior as the asymptotes of the exact solutions.

In this work we develop the semiclassical approach for relativistic
systems. We begin our analysis with the relativistic classic
consideration of the problem in the Hamilton-Jacobi formulation. Two
different versions are considered: in the first the potential is zero
component of a four-vector, and in the second the potential is
Lorentz-scalar. In the second version, the total energy of the system
is presented by analogy with non-relativistic case as sum of the
kinetic energy and the potential energy of interaction. Resulting
equation is written in covariant form of the relativistic
one-particle equation. After that, using the correspondence
principle, we derive a relativistic semiclassical wave equation
appropriate in the quasiclassical region.

It is shown that the leading-order WKB quantization condition is the
appropriate method to solve this equation for the spherically
symmetric potentials. Solution of this equation for the Coulomb and
linear potential is given. In particular, we show that unlike the
existing relativistic wave equations, solution of the semiclassical
wave equation for the scalar-like potential results in the
eigenfunctions which are regular at the spatial origin. \\

\noindent {\bf 2. Relativistic Semiclassical Wave Equation }\\

There are several wave equations to describe the motion of
relativistic particles: the Klein-Gordon equation, the Dirac
equation, the Bethe-Salpeter equation, the spinless Salpeter
equation, quasipotential equations. These equations can be solved
exactly for some spherically symmetric potentials. However, $S$-wave
solutions of these equations, for example, for the Coulomb potential
diverges at the spatial origin. Wave-function divergence at the
spatial origin is actually a general problem affecting relativistic
wave equations. For example, the solution of the Dirac equation for
the Coulomb potential for the $S$-wave states behaves as $\psi
\propto r^{\sqrt{1 -\alpha^2} -1}$ at $r\rightarrow 0$ and the
divergence of the solution to the spinless Salpeter equation is much
stronger: $\psi \propto r^{-4\alpha_s /3\pi}$ \cite{Dur}.

Essential problem of relativistic description of physical systems is
concerned with the entity of the potential. The potential is a
non-relativistic conception and it has been correctly defined in
classic mechanics as the function of the spatial variables. In
relativistic case the potential has not been correctly defined. A
serious problem is the nature of the potential: whether it is
Lorentz-vector or Lorentz-scalar or it is their mixture? This problem
is especially important in hadron physics where, for the vector-like
confining potential, there are no normalizable solutions \cite{Su}.
In this letter we investigate both of these possibilities when the
potential is the Lorentz-scalar and zero component of a 4-momentum.

To derive the wave equation with the help of the correspondence
principle, we need to know the corresponding classic equation. This
is why we consider first classic description of relativistic
particles. The classic equations are then used to derive the
corresponding semiclassical wave equations.

As known, for a relativistic particle moving in the field of the
spherically symmetric potential $V(r)$, the interaction is introduced
according to the gauge invariant principle, $p_\mu \rightarrow P_\mu
= p_\mu -eA_\mu$, where $p_\mu$ is the four-momentum of the particle
and $A_\mu$ is the external field. For zero component $P_0 = E$ this
results in the total energy of the particle in the form (hereafter we
basically use the system $\hbar = c = 1$)

\begin{equation} E = \sqrt{{\vec p}^{\,2} + m^2} + V(r),
\end{equation}
where $V(r) = eA_0$ is the static potential. Equation (1) (with the
help of the correspondence principle) results in the spinless
Salpeter equation \cite{Sal}. For the classic equation $\vec p^{\,2}=
[E-V(r)]^2 - m^2$, the same correspondence principle results in the
static Klein-Gordon equation.

Due to the success of the non-relativistic potential models, it can
be used to scrutinize the non-relativistic method. This method has
proven extremely successful for the description not only of
non-relativistic systems but also of relativistic bound states.
Potential models work much better than one would naively expect (see
Ref. \cite{Luc}). This success is somewhat puzzling in that it
persists even when the model is applied to relativistic systems.

It is generally agreed, that in nonrelativistic classic mechanics,
the total energy $E'$ of a particle of mass $m$ moving in the field
of the central-field potential $V(r)$ is given by equation $E'= T +
V(r)$, where $T$ is the kinetic energy of the particle; the total
energy $E'$ is the integral of motion. In relativistic case, the
total energy can be written in analogous form and, what is more, the
potential $V(r)$ in this case is a Lorentz-scalar.

Energy $\sqrt{\vec p^{\,2} + m^2}$ of relativistic particle in Eq.
(1) can be presented as sum of the kinetic energy $T$ and the rest
mass $m$ as $\sqrt{\vec p^{\,2} + m^2} = T + m$. Then equation (1)
takes the form

\begin{equation}  E = T + W(r),     \end{equation}
where we have introduced notation $W(r) = m + V(r)$. From
nonrelativistic point of view, $W(r)$ is the potential energy shifted
by constant value $m$ with respect to $V(r)$. On the other hand, from
relativistic point of view, $W(r)$ can be considered as a variable
mass (this topic has been discussed in series of works
\cite{Ono,Rav,Dull}). This makes sense because two ingredients of the
total energy (2), the rest mass $m$ and the potential $V(r)$, are
static quantities and they can be combined into one object, $W(r)$.
Thus we can write for the kinetic energy $T$ of the particle of the
effective mass $W(r)$, $T = E-W(r)\equiv \vec\pi^2/[E + W(r)]$, where
the squared momentum $\vec\pi^2$ is

\begin{equation} \vec\pi^2 = E^2 - W^2(r).   \end{equation}
Equation (3) has the form of the equation for a free relativistic
particle with the effective mass $W(r)$ and can be written in the
covariant form as

\begin{equation}   \pi_\mu\pi^\mu - W^2(r) = 0,    \end{equation}
where $\pi^\mu = (E,\vec\pi)$.

Now, we can derive the wave equations which correspond to the classic
equations written above. We start with the classic problem in the
Hamilton-Jacobi formulation. The static Hamilton-Jacobi equation
corresponding to the classic equation $\vec p^{\,2} = [E-V(r)]^2 -
m^2$ is

\begin{equation}
\left(\frac{\partial S_0}{\partial r}\right)^2 + \frac 1{r^2}
\left(\frac{ \partial S_0}{\partial \theta }\right)^2+\frac
1{r^2\sin^2\theta}\left( \frac{\partial
S_0}{\partial\varphi}\right)^2 = \left[E-V(r)\right]^2 - m^2,
\end{equation}
where $S_0$ is the action of the system. To obtain the corresponding
wave equation we present the wave function in the form $\tilde\psi
(\vec r) = A\,\exp\left[\frac i\hbar S(\vec r)\right]$, where $A$ is
the arbitrary constant. As was shown in Ref. \cite{Se}, in the
quasiclassical region, in representation of the wave function
$\tilde\psi(\vec r)$, the squared momentum operator has the form

\begin{equation}
\hat {\vec p^{\,2}} =
\left(-i\hbar\frac{\partial}{\partial r}\right)^2 +
\frac 1{r^2}\left(-i\hbar\frac{\partial}{\partial\theta}\right)^2 +
\frac 1{r^2\sin^2\theta}\left(-i\hbar\frac{\partial}
{\partial\varphi}\right)^2,
\end{equation}
which follows from the classic expression for $\vec p^{\,2}$ with the
help of the correspondence principle $\frac{\partial S_0}{\partial q}
\rightarrow -i\hbar\frac\partial{\partial\,q}$, $q = r,\theta
,\varphi $ \cite{Se}. Thus we obtain the wave equation ($\hbar=c=1$)

\begin{equation}
\left[\left(-i\frac{\partial}{\partial r}\right)^2 + \frac
1{r^2}\left(-i\frac{\partial}{\partial\theta}\right)^2 + \frac
1{r^2\sin^2\theta}\left(-i\frac{\partial}{\partial\varphi}\right)^2
\right]\tilde\psi(\vec r) = \left[\left(E-V(r)\right)^2 -
m^2\right]\tilde\psi(\vec r).
\end{equation}
Equation (7) is the second-order differential equation in canonical
form. This relativistic semiclassical wave equation is closely
related to the classic equation (5) and is appropriate to describe
relativistic quantum systems in the quasiclassical region. Below, we
solve this equation for some spherically symmetric potentials.

Let us consider now the classic equation (3). The corresponding
semiclassical equation can be derived analogously to the above case.
Using the operator (6), with the help of the correspondence principle
we obtain the following equation

\begin{equation}
\left[\left(-i\frac{\partial}{\partial r}\right)^2 + \frac
1{r^2}\left(-i\frac{\partial}{\partial\theta}\right)^2 + \frac
1{r^2\sin^2\theta}\left(-i\frac{\partial}{\partial\varphi}\right)^2
\right]\tilde \psi (\vec r) =
\left[E^2 - W^2(r)\right] \tilde \psi (\vec r).
\end{equation}
The correlation of the function $\tilde{\psi}(\vec r)$ with the wave
function ${\psi}(\vec r)$ in case of the spherical coordinates is
given by relation $\tilde{\psi}(\vec r) = \sqrt{det\,g_{ij}}
\psi(\vec r)$, which follows from the identity: $\int\mid \psi(\vec
r)\mid^2d^3\vec r$ $\equiv \int\mid\psi(\vec r)$ $\mid^2det\,g_{ij}
dr\,d\theta\,d\varphi $ $= 1$, where $g_{ij}$ is the metric tensor
(det$\,g_{ij}=r^2\sin\,\theta$ for the spherical coordinates). \\

\noindent {\bf 3. Solution of the Semiclassical Wave Equation }\\

Relativistic wave equations are usually solved in terms of special
functions, with the help of specially developed methods or
numerically. However, almost together with quantum mechanics, the
appropriate method to solve the wave equation has been developed; it
is general simple for all the problems, and its correct application
results in the exact energy eigenvalues for all solvable potentials.
This is the phase-integral method which is also known as the WKB
method \cite{Shi,Fro}.

The WKB method in its present form has been formulated to solve
one-dimensional two-turning-point problems. Within the framework of
the WKB method the solvable potentials mean those potentials for
which the eigenvalue problem has two turning points. However the WKB
method can be used to solve problems with more then two turning
points.

The WKB formulas can be written differently, i.e. on the real axis
\cite{Shi} and in the complex plane \cite{Fro}. For the
two-turning-point problems, the standard leading-order WKB
quantization condition is defined by

\begin{equation}
\int_{x_1}^{x_2}\sqrt{p^2(x)}dx =\pi\hbar\left(n+\frac 12\right),
\end{equation}
where $p(x)$ is the generalized momentum and $n$ is the number of
zeros of the wave function inside the interval $[x_1,x_2]$. However,
the most general form of the WKB solution and the quantization
condition can be written in the complex plane.

The WKB solution of the one-dimensional Schr\"odinger's equation is
searching for in the form $\psi(z)=a\exp\left[\frac i{\hbar}
S(z)\right]$, where $a$ is the arbitrary constant and the action of
the system, $S(z)$, is written as the expansion in powers of $\hbar$,
$S(z) = $ $S_0(z) + \hbar S_1(z) + \hbar^2S_2(z) + \dots$. In the
leading $\hbar$ approximation we have $S_0(z) = \pm\int^z p(z)dz$,
$S_1(z) = i\hbar\ln\sqrt{|p(z)|}$, where $p(z)$ is the classic
momentum. Thus, in this approximation, the WKB solution is

\begin{equation}
\psi^{WKB}(z) = \frac A{\sqrt{|p(z)|}}\exp\left[\frac i{\hbar}
\int^zp(z)dz\right].      \end{equation}

In quantum mechanics, quantum numbers are determined as number of
zeros of the wave function in the classically allowed region. In the
complex plane, number of zeros $N$ of a function $y(z)$ inside the
contour $C$ is defined by the argument's principle \cite{Kor}. For
the wave function $\psi(z)$, according to this principle we have

\begin{equation}
\oint_C \frac {\psi\prime(z)}{\psi(z)}dz = 2\pi iN,   \end{equation}
where $\psi\prime(z)$ is the derivative of the function $\psi(z)$.
Contour $C$ is chosen such that it includes cuts between the turning
points on the generalized momentum $p(z)\ge 0$ and no other
singularities.

Substituting (10) into (11) we obtain the quantization condition

\begin{equation}
\oint\sqrt{p^2(z)}dz + i\frac\hbar 2\oint\frac{p\prime(z)}{p(z)}dz  =
2\pi\hbar N.     \end{equation}
In the case, when $p(z)$ is a smooth function of the spatial variable
and equation $p^2(z)=0$ has two roots (turning points), Eq. (12)
results in the quantization condition

\begin{equation}
\oint\sqrt{p^2(z)}dz = 2\pi\hbar\left(N+\frac 12\right).
\end{equation}
In particular case $p^2(z)=const$ quantization condition has the form

\begin{equation}
\oint\sqrt{p^2(z)}dz = 2\pi\hbar N.
\end{equation}

Quasiclassical method is fruitfully used not only in quantum
mechanics but also in the relativistic theory \cite{Kang,Se1,KrSe}.
Application of the leading-order WKB quantization rule to the
nonrelativistic semiclassical wave equation \cite{Se} yields the
exact energy eigenvalues for {\em all} solvable spherically symmetric
potentials and no further Langer-like corrections are necessary.

Equations (7) and (8) are the second-order differential equations of
the Schr\"odinger type in canonical form. Important feature of these
equations is that, for the two and more turning-point problems, they
can be solved exactly by the conventional WKB method.

For the spherically symmetric potentials, Eqs. (7), (8) are
separated. Angular equation for each of them is the same,

\begin{equation}
\left[\left(-i\frac{\partial}{\partial \theta}\right)^2 + \frac
1{\sin^2\theta}\left(-i\frac{\partial}{\partial\varphi}\right)^2
\right]\tilde Y(\theta,\varphi) =
\vec M^2\tilde Y(\theta,\varphi), \end{equation}
and the corresponding radial equations are

\begin{equation} \left(-i\frac d{dr}\right)^2\tilde R(r) =
\left[(E-V(r))^2 - m^2 - \frac{\vec M^2}{r^2}\right]\tilde R(r),
\end{equation}

\begin{equation} \left(-i\frac d{dr}\right)^2\tilde R(r) =
\left[E^2 - W(r)^2 - \frac{\vec M^2}{r^2}\right]\tilde R(r).
\end{equation}

\noindent {\bf 3.1. The angular momentum eigenvalues}\\

Angular equation (15) determines the squared angular momentum
eigenvalues $\vec M^2$ which enter into the radial equations (16),
(17). The solution of equation (15) has been obtained in Ref.
\cite{Se} that gives for $\vec M^2$,

\begin{equation}  \vec M^2 = \left(l+\frac 12\right)^2,
\ \ \ l = 0, 1, 2,... \end{equation}
This same result can be obtained in the complex plane with the help
of the quantization condition (13). It is interesting from the
methodological point of view to bring here the calculation of $\vec
M^2$ in the complex plane because this calculation is shorter and
beautiful.

Separation of the angular Eq. (15) results in two one-dimensional
equations,

\begin{equation}
\left(-i\frac d{d\theta}\right)^2\tilde\Theta(\theta) =
\left(\vec M^2 - \frac{M_z^2}{\sin^2\theta}\right)
\tilde\Theta(\theta), \end{equation}

\begin{equation}
\left(-i\frac d{d\varphi}\right)^2\tilde\Phi(\varphi) =
M_z^2\tilde\Phi(\varphi), \end{equation}
The WKB quantization condition (14) which is appropriate to the
angular equation (20) [$p(\varphi) = M_z$], gives $M_z = m$, $m =
0,1,2,...$. The corresponding quasiclassical solution is

\begin{equation}
\tilde\Phi_m(\varphi) = A\,e^{im\varphi} + B\,e^{-im\varphi},
\end{equation}
where $A$ and $B$ are the arbitrary constants.

The WKB quantization condition (13) appropriate to equation (19) is

\begin{equation}
I = \oint_C\sqrt{\vec M^2 - \frac{M_z^2}{\sin^2\theta}}d\theta =
2\pi\left(n_\theta +\frac 12\right), \ \ \ n_{\theta} = 0,1,2,...
\end{equation}
To calculate integral (22) (as other hereafter) we use the method of
stereographic projection. This means that, instead of integration
about a contour $C$ enclosing the classical turning points, we
exclude the singularities outside the contour $C$, i.e., at $\theta =
0$ and $\infty $ in this particular case. Excluding these infinities
we have, for the integral (22), $I = I_0 + I_{\infty}$. Integral $I_0
= - 2\pi M_z$, and $I_{\infty}$ is calculated with the help of the
replacement $z=e^{i\theta}$ that gives $I_{\infty} = 2\pi\sqrt{\vec
M^2}$. Therefore $I = 2\pi(\sqrt{\vec M^2} - M_z)$ and we obtain
eigenvalues (18).

Equation (18) represents the squared angular momentum eigenvalues in
the quasiclassical region which are different from the known ones
$l(l+1)$ obtained from the solution of the Klein-Gordon equation.
The centrifugal term in the radial equations (16), (17) has the form
$(l+\frac 12)^2/r^2$ for {\em all} spherically symmetric potentials
$V(r)$ which is not equal to zero at $l=0$. The WKB solution
corresponding to the eigenvalues (18) has the correct asymptotic
behavior at $\theta\rightarrow 0$ and $\pi$ for all values of $l$ and
agrees with the asymptote of the spherical functions $Y_{lm}(\theta
,\varphi )$ \cite{Se}. It is important to emphasize that the squared
angular momentum eigenvalues (18) have obtained in our approach from
the solution of the angular semiclassical equation (15) in the
framework of the same quasiclassical method. \\

\noindent{\bf 3.2. The Coulomb potential $V(r)= - \alpha /r$}\\

Consider several examples to show the efficiency of the semiclassical
approach in relativistic theory. Let us deal with first the classic
problem for the Coulomb potential, i.e. the radial semiclassical
equation (16). The quantization condition (13) appropriate to Eq.
(16) with the Coulomb potential is

\begin{equation} \oint_C\sqrt{E^2 - m^2 + \frac{2\alpha E}r
 - \frac{\Lambda^2}{r^2}} = 2\pi\left(n_r+\frac 12\right),
\end{equation}
where $\Lambda^2 = (l+\frac 12)^2 - \alpha^2$ and a contour $C$
encloses the classical turning points $r_1$ and $r_2$. Using the
method of stereographic projection, we should exclude the
singularities outside the contour $C$, i.e. at $r=0$ and $\infty$.
Excluding these infinities we have, for the integral (23), $I =
2\pi(\alpha E/\sqrt{-E^2+m^2} - \Lambda)$, and for the energy
eigenvalues this gives

\begin{equation} E_n = \frac m{\sqrt{1 + \frac{\alpha^2}{\left(n_r +
\frac 12 + \Lambda\right)^2}}}.  \end{equation}
Equation (24) coincides with the one obtained from the solution of
the Klein-Gordon equation with the Coulomb potential by known exact
method.

Now, let us deal with the semiclassical wave equation (17). The
quantization condition (13) appropriate to Eq. (17) with the
scalar-like Coulomb potential is

\begin{equation}
\oint_C\sqrt{E^2 -\left(m -\frac{\alpha} r\right)^2 -
\frac{(l+\frac 12)^2}{r^2}}dr = 2\pi\left(n_r +\frac 12\right).
\end{equation}
The phase-space integral (25) is calculated analogously to the above
case in closed form that results in the energy eigenvalues

\begin{equation} E_n = m\sqrt{-\frac{\alpha^2}{\left[n_r +
\frac 12 + \sqrt{(l+\frac 12)^2 + \alpha^2}\right]^2} + 1}.
\end{equation}
Formula (26) can be written in apparent relativistic form, $E_n^2 =
m^2+p_n^2$, where $p_n$ is given by equation

\begin{equation} p_n = \frac{i\alpha m}{n_r+\frac 12 +
\sqrt{(l+\frac 12)^2 + \alpha^2}}.      \end{equation}

The radial quasiclassical eigenfunctions inside the region of classic
motion $[r_1,r_2]$ are written with the help of the Eq. (10)

\begin{equation} \tilde R^{WKB}(r) = \frac A{\sqrt{|p(r)|}}
\cos\left(\int_{r_1}^r\sqrt{p^2(r)}dr - \frac{\pi}4\right).
\end{equation}
So far, as $|p(r)|\simeq\frac\Lambda r$ at $r\rightarrow 0$, this
gives, for the WKB solution in the representation of the wave
function $\psi(\vec r)$, $R(r) =\tilde R^{WKB}(r)/r\propto
r^{\Lambda}/\sqrt r$. Because $\Lambda =\sqrt{(l+\frac 12)^2 +
\alpha^2}$, for the $S$-wave states, this results in $R(r)\propto
r^{\alpha}$, i.e. $R(r)$ is regular at small $r$. We obtain this
result because the potential in Eq. (17) is {\em scalar-like}.  Note
that in the case of the Klein-Gordon equation the radial $S$-wave
function $R(r)$ diverges at $r\rightarrow 0$, i.e.  $R(r)\propto
r^{-\alpha}$.

The quasiclassical solution (28) can be written in the form of a
standing wave. For this note that the hermiticity of the operator (6)
implies the adiabatically slow alteration of the derivatives
$\frac{\partial S_0}{\partial q}$, $q = r,\theta,\varphi$, i.e.
$\frac{\partial S_0}{\partial q}\simeq const$ \cite{Se}. This
condition anticipates the final result, i.e. discrete constant
eigenvalues $p_n$ of the operator (6). Integration of equation
$\frac{\partial S_0}{\partial r}\simeq const$ gives, for the action
$S_0$, $S_0(r) = |p_n | r + const$, where $p_n$ is given by Eq. (27)
(for this problem). In the region of classic motion, where $p(r)>0$,
the solution has the form of a standing wave

\begin{equation}
\tilde R_n(r) =C\cos\left(|p_n| r -\chi_1 - \frac{\pi}4\right),
\end{equation}
where $\chi_1$ is the value of the phase integral (25) at the turning
point $r_1$. Analogous solutions can be written for other
one-dimensional equations obtained after separating of Eq. (8).

Thus we obtain two results for the Coulomb potential by the WKB
method: the known exact eigenvalues for spinless particles (24) which
coincide with those obtained from the solution of the Klein-Gordon
equation and the other ones (26) obtained from the solution of the
semiclassical equation (8) with the scalar-like Coulomb potential.
We compare two relativistic formulas (24) and (26) for the energy
eigenvalues of an electron moving in the field of the static
spherically symmetric Coulomb potential in Table 1.

\newpage
\medskip
\begin{center}Table 1. {\small Energy levels of the hydrogen atom }
\end{center}
\begin{center}
\begin{tabular}{lllll} \hline \hline
$ l$ & $ n_r$ & $\ \ \ \ \ \ \ E_n^{NR}$ & $\ \ \ \ \ \ \ \ \ \
E_n^{KG}$ & $\ \ \ \ \ \ \ \ \ \ E_n^{SC}\ \ \ $          \\
\hline
\end{tabular}
\end{center}

\begin{center}
\begin{tabular}{lllll}
$\ 0$ & $\ 0$ &  $-13.6155700$& $-13.6164800$ &  $-13.6143000$ \\
$\ 1$ & $\ 0$ &\ $-3.4038930$ &\ $-3.4039190$ &\ $-3.4038440$ \\
$\ 0$ & $\ 1$ &\ $-3.4038930$ &\ $-3.4040400$ &\ $-3.4037230$ \\
$\ 1$ & $\ 1$ &\ $-1.5128410$ &\ $-1.5128520$ &\ $-1.5128250$ \\
$\ 2$ & $\ 1$ &\ $-0.8509732$ &\ $-0.8509756$ &\ $-0.8509693$ \\
$\ 2$ & $\ 2$ &\ $-0.5446228$ &\ $-0.5446243$ &\ $-0.5446208$ \\
$\ 3$ & $\ 2$ &\ $-0.3782103$ &\ $-0.3782109$ &\ $-0.3782095$ \\
$\ 4$ & $\ 2$ &\ $-0.2778688$ &\ $-0.2778690$ &\ $-0.2778684$ \\
$\ 4$ & $\ 3$ &\ $-0.2127433$ &\ $-0.2127435$ &\ $-0.2127430$ \\
\hline \hline
\end{tabular} \end{center}
In Table 1 $E_n^{NR}$ denotes the eigenenergies obtained from
solution of the Schr\"odinger equation and $E_n^{KG}$ are the energy
eigenvalues obtained from the Klein-Gordon equation, $l$ and $n_r$
are the orbit and radial quantum number, respectively. Eigenenergies
$E_n^{SC}$ have been calculated with the help of Eq. (26).

As is well known, the energy spectrum resulting from the Klein-Gordon
equation with the Coulomb potential contains a fine-structure effect
giving an over-estimation of the fine-structure level spreading,
while the result from non-relativistic Schr\"odinger equation does
not \cite{Shi}. In our case, the eigenvalues $E_n^{SC}$ are lower
than $E_n^{NR}$ and $E_n^{KG}$, hereby $E_n^{SC}$ are closer to
non-relativistic ones. Fine-structure term which follows from
$E_n^{SC}$ is

\begin{equation}
\Delta E_n^{SC} = \frac{m\alpha^4}{2n^3}\left(\frac 1{l+\frac 12} -
\frac 1{4n}\right),
\end{equation}
where $n=n_r+l+1$. This term is different from the corresponding
correction which follows from $E_n^{KG}$ by sign and the second
term.\\

\noindent{ \bf 3.3. The linear potential $V(r)=\kappa r$}\\

The linear potential is one of the special interest in hadron
physics.  In relativistic case, this potential results in the linear
Regge trajectories which observed at experiment. In relativistic
potential models of quarkonia based on a Dirac-type equation with a
local potential is a sharp distinction between a linear potential $V$
which is vector-like and one which is scalar-like. There are
normalizable solutions for a scalar-like $V$ but not for a
vector-like $V$ \cite{Su}. No other problems arise and no
difficulties encountered with the numerical solution if the
confinement potential is purely {\em scalar-like}. It was shown in
many works that the effective interaction has to be scalar in order
to confine particles inside the hadrons (see, for example, Refs.
\cite{Su,SC}).

In relativistic theory the linear potential represents the so-called
"insoluble" problem with four turning points. This problem is
insoluble also from the viewpoint of the conventional WKB method.
However, this problem can be solved by the WKB method in the complex
plane. In the semiclassical equation (17), the potential is a
scalar-like. Consider this equation for a two-particle system of
equal masses interacting by means of the linear potential. The radial
semiclassical equation (17) is then:

\begin{equation}
\left(-i\frac d{dr}\right)^2\tilde R(r) =\left[\frac{E^2}4 -\left(m
+\kappa r\right)^2 -\frac{(l+\frac 12)^2}{r^2}\right]\tilde R(r),
\end{equation}
From four turning points of this problem, two of them, $r_1$ and
$r_2$, lie in nonphysical region $r<0$, and the other two, $r_3$ and
$r_4$ lie in the physical region $r>0$. To calculate the phase-space
integral in the complex plane we chose a contour $C$ enclosing the
cuts (and, therefore, zeros of the wave function) at $r>0$ and $r<0$
between the turning points $r_1$, $r_2$ and $r_3$, $r_4$,
respectively. Thus, the quantization condition appropriate to Eq.
(31) is:

\begin{equation} I =\oint_C\sqrt{\frac{E^2}4 - (m +\kappa r)^2 -
\frac{(l+\frac 12)^2}{r^2}}dr = 4\pi\left(n_r+\frac 12\right),
\end{equation}
where $n_r$ denotes zeros of the wave function at $r>0$.

Outside the contour $C$, the problem has two singularities, i.e. at
$r=0$ and $\infty$. Using the method of stereographic projection, we
have, for the integral (32), $I = I_0 + I_{\infty}$, where $I_0 =
-2\pi(l+\frac 12)$ and the integral $I_{\infty}$ is calculated with
the help of the replacement of variable, i.e. $z=\frac 1r$, that
gives $I_{\infty} = 2\pi E^2/(8\kappa)$. Therefore, for $E_n^2$, we
obtain

\begin{equation} E_n^2 = 8\kappa\left(2n_r + l + \frac 32\right),
\end{equation}
which is very similar to that of a harmonic oscillator-type
Hamiltonian.

It is an experimental fact that the dependence $E_n^2(l)$ is linear
for light mesons (linear Regge trajectories \cite{Coll}). However, at
present, the best way to reproduce the experimental masses of
particles is to rescale the entire spectrum assuming that the masses
$M$ of the mesons are expressed by the relation \cite{SC}

\begin{equation}     M_n^2 = E_n^2 - C^2,     \end{equation}
where $E_n$ is given by (33) and $C$ is a constant energy (free
parameter). Relation (34) is used to shift the spectra and appears as
a means to simulate the effects of unknown structure approximately.

However, the required shift of the spectra naturally follows from the
solution of the semiclassical equation (17). To describe the Regge
trajectories observed at experiment one needs to take into account
the week coupling effect. In this case the potential includes the
Coulomb-like term, $-\alpha_s/r$, i.e. the scalar-like potential is
$V(r) = -\alpha_s/r + \kappa r$ (the funnel potential). For light
mesons one may expect that the $q\bar q$ bound states will feel only
the linear part of the potential which gives the main contribution to
the binding energy. We thus assume that the Coulomb term can be
considered as a small perturbation. Then we obtain instead of (33)

\begin{equation}
E_n^2 = 8\kappa\left(2n_r +l -\alpha_s +\frac 32\right).
\end{equation}

Equation (35) does not require any additional free parameter. We
obtain the necessary shift with the help of the term
$-8\kappa\alpha_s$ which is the result of interference of the Coulomb
and linear terms of the interquark potential. Note that we obtain the
correct sign (minus) for this term only in the case of the {\em
scalar-like} potential. Equation (35) gives a good description of
mass of light vector mesons for the typical value $\kappa\simeq
0.14\,GeV^2$, i.e. reproduces the linear dependence of angular
momentum $l$ on $E^2$ that is in agreement with the experimental data
for light mesons \cite{Coll}. \\

\noindent {\bf 4. Conclusion }\\

In this letter, we have developed the semiclassical approach to the
description of relativistic systems. In our analysis we have started
with the relativistic classic consideration of the problem in the
Hamilton-Jacobi formulation. Two different possibilities have been
considered: in the first the potential is the zeroth component of a
four-vector and in the second, the potential is the Lorentz-scalar.

The main result is the derivation of the relativistic semiclassical
wave equation [(7) for the vector-like potential and (8) for the
scalar-like potential] appropriate in the quasiclassical region. This
equation is the second-order differential equation in canonical form
(without first derivatives). In case of the scalar-like potential,
the resulting equation has been written in covariant form.  Unlike
the Klein-Gordon equation, the semiclassical one (in case of the
vector-like potential) results in another integral of motion, i.e.
the squared angular momentum $\vec M^2=(l+\frac 12)^2\hbar^2$.  This
is same squared angular momentum we have obtained also for the
semiclassical equation with the scalar-like potential.

The squared angular momentum eigenvalues $\vec M^2=(l+\frac
12)^2\hbar^2$ have been obtained in our approach in a natural way
from the solution of the angular semiclassical equation (15) in the
framework of the same WKB method that gives the justification of the
Langer correction as that to the squared angular momentum
eigenvalues. This means that the centrifugal term in the radial
semiclassical equation has the form $(l+\frac 12)^2/r^2$ for {\em
any} spherically symmetric potential $V(r)$.

Important feature of the semiclassical wave equation is that it can
be solved by the WKB method. We have shown that the standard
leading-order WKB quantization condition reproduces the exact
eigenvalues for two-turning-point problems. Besides, this equation
can be solved by the WKB method for more then two turning-point
problems.

To show efficiency of the semiclassical approach in relativistic
theory we have considered several examples. In particular, we have
obtained two results for the Coulomb potential by the WKB method: the
known exact result for spinless particles (24) which coincides with
one obtained from solution of the Klein-Gordon equation and another
result (26) obtained from the solution of the semiclassical equation
(8) for the scalar-like Coulomb potential. We have shown that, unlike
the known relativistic wave equations for the Coulomb potential, the
semiclassical one for the scalar-like Coulomb potential has the
regular solution at the spatial origin. Conception of the scalar-like
potential is especially important in hadron physics. The solution of
equation (8) for light $q\bar q$ states with the funnel potential has
allowed us to reproduce the linear dependence $E^2(l)$ and obtain
apparent form of the shift parameter $C^2$, $C^2 = -8\alpha\kappa$.

{\it Acknowledgment.} This work was supported in part by the
Belarusian Fund for Fundamental Researches.

\end{document}